\documentclass[conference]{IEEEtran}
\usepackage{cite}
\usepackage{amsmath,amssymb,amsfonts}
\usepackage{graphicx}
\usepackage{textcomp}
\usepackage{xcolor}
\def\BibTeX{{\rm B\kern-.05em{\sc i\kern-.025em b}\kern-.08em
    T\kern-.1667em\lower.7ex\hbox{E}\kern-.125emX}}

\usepackage{delarray}
\usepackage{mdwmath}
\usepackage{mdwtab}
\usepackage{eqparbox}
\usepackage{stfloats}
\usepackage{subfloat}
\usepackage{floatrow}
\usepackage[none]{hyphenat}
\usepackage{amsbsy}
\usepackage{algorithm}
\usepackage[noend]{algpseudocode}
\usepackage{listings}
\usepackage{array,multirow}
\usepackage{nicefrac}
\usepackage{mathtools}
\usepackage{verbatim}

\usepackage{enumitem}
\setlist{nosep, leftmargin=14pt}

\usepackage[italicdiff]{physics}

\usepackage{units}
\usepackage{mathtools}

\usepackage{multirow}
\usepackage{url}

\usepackage[position=bottom]{subfig}

\usepackage{tikz}

\newcommand\copyrighttext{%
  \footnotesize 979-8-3195-2276-4/26/\$31.00~\copyright2026 IEEE}
\newcommand\copyrightnotice{%
\begin{tikzpicture}[remember picture,overlay]
    \node[anchor=south west,xshift=1in,yshift=0.5in] at (current page.south west) {\copyrighttext};
\end{tikzpicture}%
}

\begin{document}

\flushbottom
\enlargethispage{-\baselineskip}

\title{Doppler Tomography Using Rydberg Sensors\\ }

\author{\IEEEauthorblockN{Peter Vouras$^1$, Bariscan Yonel$^2$, Alexandra Artusio-Glimpse$^3$}
{\textit{$^1$U.S. Department of Defense, $^2$University at Albany, $^3$National Institute of Standards and Technology}} }

\maketitle
\copyrightnotice
\vspace{-\topskip} 

\begin{abstract}
Novel sensors that leverage the quantum properties of atoms for measuring propagating electromagnetic fields are becoming increasingly practical for a variety of applications.  These sensors rely on the phenomenon of electromagnetically induced transparency (EIT), which is induced in a confined vapor of alkali atoms when the atoms are excited to a high-energy quantum state, known as a Rydberg state, with multiple resonant optical fields.  In this state, the atoms are highly sensitive to electromagnetic radiation and yield a measurement output proportional to the magnitude of an impinging electric field when resonant with a Rydberg-Rydberg transition.  In this paper, we consider the use of Rydberg sensors for a tomographic imaging application through a set of modeled system dynamics.  Our contribution includes a novel method for placing nulls in the image by modulating the radiated local oscillator (LO) that is used to recover phase information from the received signal.  We also present an algorithm for deblurring the image.
\end{abstract}


\section{Introduction}

In recent years, quantum sensors based on Rydberg atom technology have made remarkable progress towards becoming viable alternatives to conventional antennas for detecting RF signals~\cite{R21,R22,R23,R24,R25,R26,R27,R28,R29,R30,R31}.  In the most sensitive operating regime, Rydberg sensors leverage the electromagnetically induced transparency (EIT) phenomenon to measure the amplitude of electromagnetic waves propagating across a cloud of alkali metal atoms trapped in a glass vapor cell.

In the EIT configuration, a near-infrared laser beam (known as the probe laser) propagates through a glass vapor cell containing the trapped alkali atoms.  The frequency of the laser is resonant with the D2-line~\cite{Steck2025Rb85} of rubidium or cesium, most commonly.  The probe laser is absorbed by the atoms so the light that reaches a photodetector located at the opposite end of the cell is severely attenuated.  A second laser at visible wavelengths (known as the coupling laser) counter propagates along the same path as the probe laser through the vapor cell.  The coupling laser creates a narrow transparency region in the atoms' absorption spectrum as a function of probe or coupling frequency that allows more light from the probe laser to reach the photodetector, i.e., induced transparency.  The coupling laser additionally excites the valence electron of the alakli atoms to a high-lying Rydberg state where the energy separation to neighboring states is in the radio-frequency (RF) domain. If these Rydberg atoms are exposed to a RF field, then the EIT transparency window splits into two peaks in a phenomenon known as Autler-Townes (AT) splitting.

The advantage of exploiting AT splitting for RF field measurements is that the frequency difference ${\Delta{f}}$ between the two peaks in the absorption spectrum depends only on fundamental constants of nature and the magnitude of the electric field, ${E}$, as in~\cite{R30}
\begin{equation}
    \label{E:eqn1}
    \Delta{f} = \frac{p}{2{\pi}\hbar}\lvert{E}\rvert.
\end{equation}
Here, ${p}$ is the electric dipole moment of the atoms and ${\hbar}$ is the reduced Planck constant~\cite{CODATA2018}.  Thus, electric field measurements using Rydberg sensors do not require calibration since the quantum states of atoms do not drift.

Rydberg sensors translate the field amplitude of incoming electromagnetic radiation to the modulated power of the probe laser beam being detected by a photodetector.  This power measurement does not directly support the heterodyne reception of RF signals.  To recover the phase information embedded in the received RF signal, it is necessary to employ an interference technique by which the Rydberg sensors are simultaneously radiated by a local oscillator (LO) signal.  If the LO signal is at a frequency ${\omega_{LO}}$ close to the frequency ${\omega_{0}}$ of the impinging signal, then the modulated output of the probe laser will oscillate at the frequency ${\omega_{LO}-\omega_{0}}$ with the original signal phase intact~\cite{R2, R3}.  This phase can now be retrieved after downconverting the beat signal to baseband for analog-to-digital sampling.  In this paper, we show how to apply a continuous phase offset to the LO signal so that it creates point nulls in the output tomographic image.

\section{Measurement Configuration for Doppler Tomography}
Coherent Doppler tomography is a technique used to image the scattering centers on a rotating object and has applications ranging from medical ultrasound to planet observations.  Doppler tomography collects data in the transform domain and is well suited for narrowband systems as it measures the Doppler shifts from a rotating object using a single radiated frequency~\cite{R1,R4,R5,R6,R7,R8,R9,R10,R11,R12,R13,R14,R15,R16,R17,R18,R19,R20}.  As the object rotates, scattering centers along a line of constant cross range have the same radial velocity component towards the receiving antenna and create a Doppler shift in the received signal.  For a fixed time instant, the instantaneous amplitudes along different Doppler frequencies can be interpreted as a tomographic projection.  As the object continues to spin, additional projections are measured for each rotation angle.  Over the course of a full revolution, the measured data populate k-space and can be coherently processed to yield an image of the object~\cite{R1}.

\begin{figure}[h!]
  \centering
  \centerline{\includegraphics[width=0.48\columnwidth]{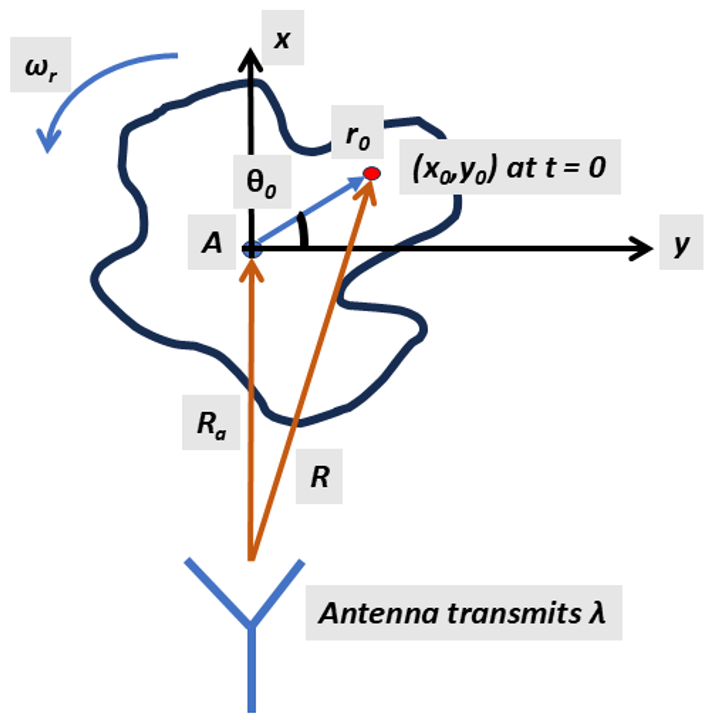}}
\caption{Measurement configuration for coherent Doppler tomography showing an object uniformly rotating about an axis through the point ${A}$ while being illuminated by an unmodulated carrier at a single frequency.}
\label{fig-1}
\end{figure}

Consider the measurement configuration shown in Fig. \ref{fig-1}.  Here, an object with a scattering center located at coordinates ${(x_0,y_0,z_0)}$ at time ${t=0}$ is rotating about the point ${A}$ with a constant angular velocity ${\omega_r}$.  ${R_a}$ denotes the distance from the antenna to ${A}$ and ${R}$ is the time varying range to the scatterer. Throughout this paper, we assume that ${R_a{\gg}2D^2/{\lambda}}$, where $\lambda$ is the carrier wavelength of the transmit antenna and ${D}$ is its largest dimension.  The parameter ${\theta_0}$ is the initial angle of the scatterer with respect to the y-axis and ${r_0}$ is the radial distance of the scatterer from ${A}$.

The time varying range between the antenna and the scatterer is given by~\cite{R35},
\begin{equation}
    R(t) = [r_0^2 + R_a^2 +2R_ar_0\sin(\theta_0 + \omega_rt) + z_o^2]^{0.5}
\end{equation}
and assuming ${R_a >> r_0, z_0}$, then ${R(t)}$ can be approximated by,
\begin{equation}
    R(t) \approx R_a + x_0 \sin{(\omega_rt) + y_0\cos{(\omega_rt)}}.
\end{equation}
The time-varying phase due to the propagation distance is now,
\begin{equation}
    \psi(t) = \frac{4{\pi}R(t)}{\lambda}
\end{equation}
and the received signal can be written as,
\begin{equation}
    s(\Theta) = s(\omega_rt) = a(t)e^{-\textrm{j}\psi(t)} = a(t)e^{-\textrm{j}4{\pi}R(t)/{\lambda}},
\end{equation}
where ${a(t)}$ denotes the signal amplitude and ${\Theta = \omega_rt}.$
The Doppler shift, ${f_d}$, of the received signal is,
\begin{align}
    f_d &= \frac{1}{2\pi}\dv{t}\psi(t) = \frac{2}{\lambda}\dv{t}R(t)   \\   \nonumber
    &= \frac{2x_o{\omega_r}}{\lambda}\cos(\omega_rt) - \frac{2y_o{\omega_r}}{\lambda}\sin{(\omega_rt)}.
\end{align}

\section{Image Reconstruction Algorithm}
Each rotation angle ${\Theta}$ corresponds to spatial frequency coordinates ${(u,v)}$ in k-space through the equations,
\begin{equation}
    u = \frac{2}{\lambda}\sin(\Theta), \quad v = -\frac{2}{\lambda}\cos(\Theta).
\end{equation}
The delta function appears because if the spatial spectrum is viewed in polar coordinates, then all the spatial frequencies appear on a circle with radius ${2/{\lambda}}$.  As a scatterer rotates through a complete revolution, the received signal yields the spatial frequency spectrum ${s(u,v)}$.  To generate an image of the object reflectivity function ${g(x,y)}$ in Cartesian coordinates, the spatial spectrum is inverted using an inverse Fourier transform as in,
\begin{equation}
    g(x,y) = \int \int s(u,v) e^{-\textrm{j}2{\pi}(ux+vy)}dudv.
\end{equation}
To describe the reflectivity function ${g(x,y)}$ in polar image coordinates ${g(r,\nu)}$, let
\begin{equation}
    x = r\cos{\nu}, \quad y = r\sin{\nu}
\end{equation}
and the differential area ${du\text{ }dv}$ becomes,
\begin{equation}
    du \text{ } dv \rightarrow \rho \text{ } d{\rho} \text{ } d{\Theta}.
\end{equation}
Then,
\begin{align}
    &g(r,\nu) = \\ \nonumber
    &\int_{0}^{\infty}\int_{0}^{2\pi}s(\Theta)\delta(\rho-2/{\lambda})e^{-\textrm{j}(4{\pi}r/{\lambda})[\cos\nu\sin{\Theta} - \sin\nu\cos\Theta]}{\rho}d{\rho}d{\Theta}  \\ \nonumber
    &= \int_{0}^{\infty}\int_{0}^{2\pi}s(\Theta)\delta(\rho-2/{\lambda})e^{-\textrm{j}(4{\pi}r/{\lambda})\sin({\Theta - \nu})}{\rho}d{\rho}d{\Theta}  \\ \nonumber
    &= \frac{2}{\lambda}\int_{0}^{2\pi}s(\Theta)e^{-\textrm{j}(4{\pi}r/{\lambda})\sin({\Theta - \nu})}d{\Theta}.
\end{align}
On a discrete grid of polar image coordinates ${(r_m, \nu_n)}$,
\begin{equation}
    g(r_m,\nu_n) \approx \frac{2}{\lambda} \sum_{k=1}^{P} s(\Theta_k)e^{-\textrm{j}(4{\pi}r_m/{\lambda})\sin(\Theta_k-\nu_{n})},
\end{equation}
where ${P}$ is the total number of time samples.

\section{Computing a Phase Offset for Image Nulls}
The problem considered in this section is to create a point null in the image such that ${g(r_m,\nu_{n})=0}$ or ${g(x_m,y_n)=0}$.  The null is created by using the radiated LO to impart a continuous phase offset to the received signal.  The received signal with a continuous phase modulation ${\phi(t)}$ applied is~\cite{R34},
\begin{equation}
    \tilde{s}(t) = a(t)e^{\textrm{j}[\psi(t)+\phi(t)]}.
\end{equation}
Write the reflectivity function as,
\begin{align}
    &g(r_m,\nu_n, \psi,\phi,a) = \\ \nonumber
    &\frac{2}{\lambda}\int_{0}^{2\pi}a(\Theta)e^{\textrm{j}\psi(\Theta)}e^{\textrm{j}\phi(\Theta)}e^{-\textrm{j}(4{\pi}r_m/{\lambda})\sin(\Theta-\nu_{n})}d\Theta.
\end{align}
Extracting linear terms from the Taylor series yields,
\begin{align}
    &g(r_m,\nu_n, \psi,\phi,a) \approx g(r_m,\nu_n, \psi, 0,a) +  \\  \nonumber
    &\frac{2}{\lambda}\int_{0}^{2\pi}
    \left[\frac{\partial}{\partial{\phi}}g(r_m,\nu_n, \psi,\phi,a)\right]
    \Bigg|_{\phi=0}\phi(\Theta)d\Theta \\
    & = 0.
\end{align}
The functional derivative in brackets is equal to,
\begin{align}
    &\frac{\partial}{\partial{\phi}}g(r_m,\nu_n, \psi,\phi,a)
    \Bigg|_{\phi=0} = \\  \nonumber
    &\textrm{j}a(\Theta)e^{\textrm{j}\psi(\Theta)}e^{-\textrm{j}\phi(\Theta)}e^{-\textrm{j}(4{\pi}r_m/{\lambda})\sin(\Theta-\nu_n)}\Bigg|_{\phi=0}  \\  \nonumber
    &=\textrm{j}a(\Theta)e^{\textrm{j}\psi(\Theta)}e^{-\textrm{j}(4{\pi}r_m/{\lambda})\sin(\Theta-\nu_n)}.
\end{align}
Thus,
\begin{align}
    &g(r_m,\nu_n, \psi,\phi,a) \approx  \\ \nonumber
    &\frac{2}{\lambda}\int_{0}^{2\pi}a(\Theta)e^{\textrm{j}\psi(\Theta)}e^{-\textrm{j}(4{\pi}r_m/{\lambda})\sin(\Theta-\nu_{n})}d\Theta  \\ \nonumber
    &+ \frac{2\textrm{j}}{\lambda}\int_{0}^{2\pi}a(\Theta)e^{\textrm{j}\psi(\Theta)}e^{-\textrm{j}(4{\pi}r_m/{\lambda})\sin(\Theta-\nu_{n})}{\phi(\Theta)}d\Theta  \\  \nonumber
    &= 0.
\end{align}
Now, define the inner product,
\begin{equation}
    \langle{g,h\rangle = \int_{0}^{2{\pi}}}g(\Theta)h(\Theta)d\Theta.
\end{equation}
Then,
\begin{align}
    &g(r_m,\nu_n,\psi,\phi,a) = 0 = \\ \nonumber 
    &\frac{2}{\lambda}\left\langle{a(\Theta)e^{\textrm{j}\psi(\Theta)}e^{-\textrm{j}(4{\pi}r_m/{\lambda})\sin(\Theta-\nu_{n})}},1\right\rangle  +  \\ \nonumber
    &\textrm{j}\frac{2}{\lambda}\left\langle{a(\Theta)e^{\textrm{j}\psi(\Theta)}e^{-\textrm{j}(4{\pi}r_m/{\lambda})\sin(\Theta-\nu_{n})}},\phi(\Theta)\right\rangle. 
\end{align}
This implies,
\begin{align}
    &\textrm{j}\left\langle{a(\Theta)e^{\textrm{j}\psi(\Theta)}e^{-\textrm{j}(4{\pi}r_m/{\lambda})\sin(\Theta-\nu_{n})}},1\right\rangle  =  \\ \nonumber
    &\left\langle{a(\Theta)e^{\textrm{j}\psi(\Theta)}e^{-\textrm{j}(4{\pi}r_m/{\lambda})\sin(\Theta-\nu_{n})}},\phi(\Theta)\right\rangle. 
\end{align}
Let,
\begin{align}
    &\mathbf{w} = a(\Theta)e^{\textrm{j}\psi(\Theta)} \times \\ \nonumber
    & \left[ \begin{array}{ccc}
    e^{-\textrm{j}(4{\pi}r_{m_1}/{\lambda})\sin(\Theta-\nu_{n_1})} & \ldots &  
    e^{-\textrm{j}(4{\pi}r_{m_K}/{\lambda})\sin(\Theta-\nu_{n_K})} \end{array}  \right],
\end{align}
where ${K}$ is the total number of desired nulls.
Then,
\begin{equation}
    \textrm{j}\langle\mathbf{w},1\rangle = \langle\mathbf{w},\phi(\Theta)\rangle.
\end{equation}
Define,
\begin{equation}
    \mathbf{c} = \Re({\mathbf{w}}), \quad \mathbf{d} = \Im({\mathbf{w}}).
\end{equation}
Then,
\begin{equation}
    \langle \left[ -\mathbf{d} \quad \mathbf{c} \right] , 1\rangle] = \langle \left[ \mathbf{c} \quad \mathbf{d} \right], \phi({\Theta}) \rangle.
\end{equation}
Finally, the desired phase offset function ${\hat{\phi}({\Theta})}$ is given by,
\begin{equation}
    \hat{\phi}({\Theta}) = \left[ \mathbf{c} \quad \mathbf{d} \right] \langle \left[ \mathbf{c} \quad \mathbf{d} \right], \left[ \mathbf{c} \quad \mathbf{d} \right] \rangle^{-1} \langle \left[ -\mathbf{d} \quad \mathbf{c}\right], 1 \rangle.
\end{equation}

\section{Blurring}
In simple cases, the narrowband response of the Rydberg sensor~\cite{PhysRevA.111.033115} can be modeled as a linear blurring kernel, ${z(\Theta)}$, which convolves the received signal to yield the distorted signal,
\begin{equation}
    \bar{s}(\Theta) = s(\Theta) \ast z(\Theta) + n(\Theta).
\end{equation}
The result of coherently processing ${\bar{s}(\Theta)}$ will be a blurred image~\cite{R32,R33}.  In this section we describe an approach to deblur the image assuming the kernel ${z(\Theta)}$ is known.

To start, write the distorted signal at the output of the Rydberg sensor using a linear model as
\begin{equation}
    \mathbf{\bar{s}} = \mathbf{Z}\mathbf{s} + \mathbf{n},
\end{equation}
where ${\bar{\mathbf{s}} = \left[ \bar{s}(0) \quad \bar{s}(1) \ldots \bar{s}(L+P-2) \right]}$ is the vector of blurred signal samples, ${\mathbf{s}}$ is the ${P\times1}$ vector of the ideal signal samples without blurring, ${\mathbf{n}}$ is the vector of ${L+P-1}$ additive noise samples, and the convolution operation is represented using the ${(L+P-1) \times P}$ matrix ${\mathbf{Z}}$ constructed from ${L}$ complex samples of the blurring kernel ${\mathbf{z}}$ as,
\begin{equation}
    \mathbf{Z} = \left[ \begin{array}{cccc} z_0 & 0 & \cdots & 0 \\
    \vdots & z_0 & \vdots & 0 \\ 
    z_{L-1} & \vdots & \ddots & \vdots \\
    \vdots & z_{L-1} & \ddots & z_0 \\
    0 & \vdots & \ddots & \vdots \\
    0 & \cdots & 0 & z_{L-1} 
    \end{array} \right].
\end{equation}
Then, the estimate of the deblurred signal is given by,
\begin{equation}
    \hat{\mathbf{s}} = (\mathbf{Z}^H\mathbf{Z})^{-1}\mathbf{Z}^H{\bar{\mathbf{s}}}.
\end{equation}

\section{Simulation Results}
\begin{figure}[htb]
    \centering
    \subfloat{\includegraphics[width=0.48\columnwidth]{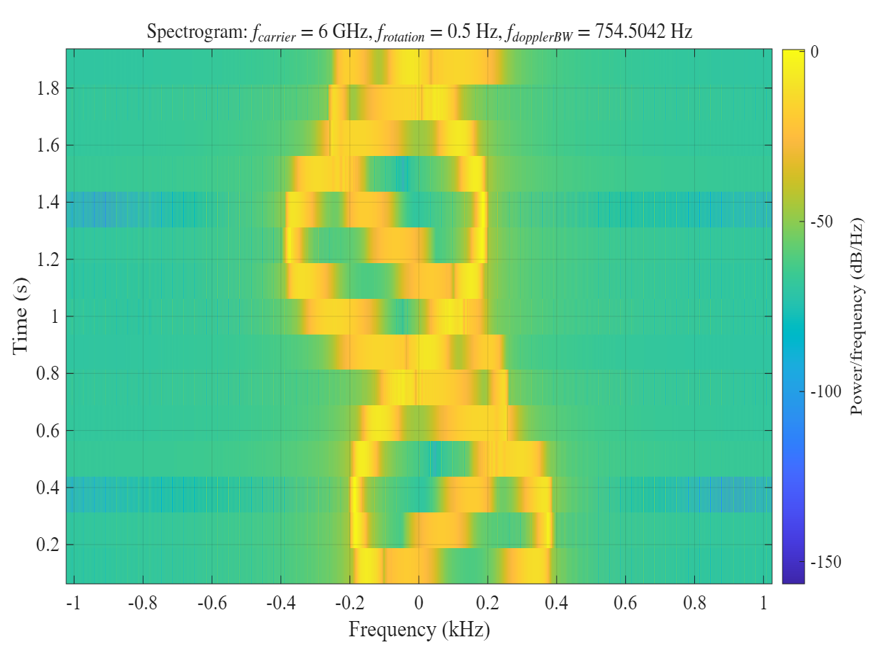}}\hspace{1pt}
    \subfloat{\includegraphics[width=0.48\columnwidth]{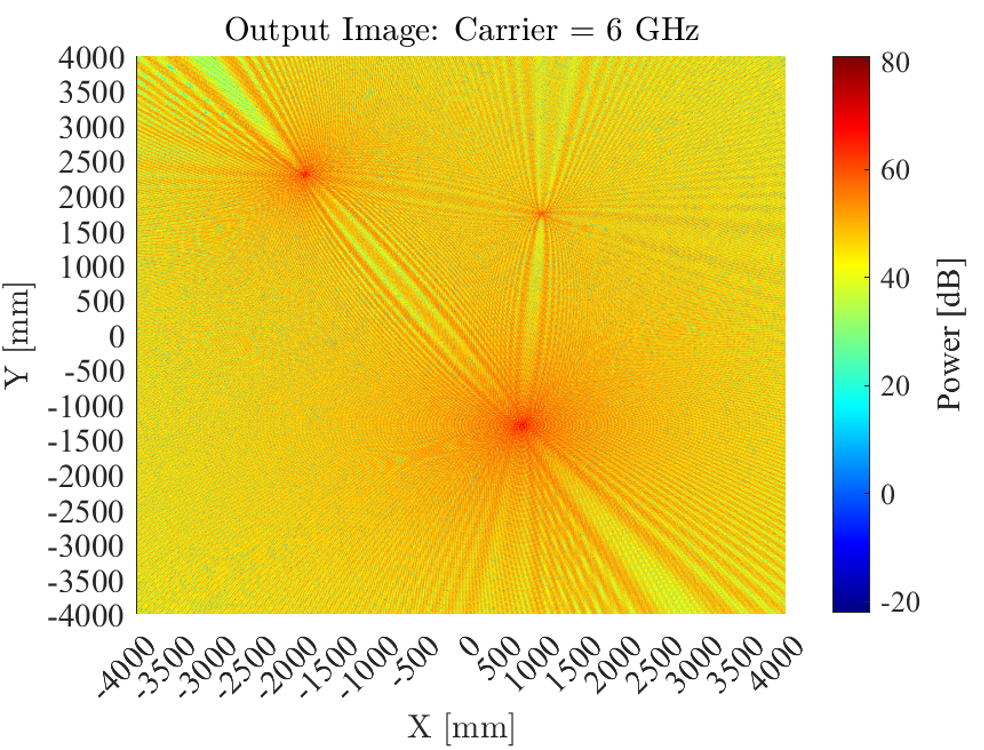}}\hspace{1pt}
    \vspace{-6pt}
	\caption{(Left) Spectrogram of received signal.  (Right) Tomographic image showing three strong scatterers in the absence of noise.}
\label{fig:Fig-2}
\end{figure}

This section presents simulation results that demonstrate the efficacy of the proposed nulling and deblurring algorithms.  For the first simulation scenario, the carrier frequency is 6~GHz, ${\omega_r} = 0.5$ rev/sec or ${\pi}$ rad/sec, and ${R_a=60}$~m.  There are 3 scatterers in the scene located at the radial offsets ${r_0 = 3}~\mathrm{m}, 2~\mathrm{m}, \mathrm{ and } 1.5~\mathrm{m}$, and the initial angular offsets are ${\theta_0 = 130^{\circ},60^{\circ},\mathrm{ and }300^{\circ}}$.  The amplitudes of the scatterers are ${2~\mathrm{V},1~\mathrm{V},}$ and ${3~\mathrm{V}}$.  Fig.~\ref{fig:Fig-2} shows a spectrogram of the received signal and the time varying Doppler shifts are clear.  Also, in Fig. \ref{fig:Fig-2} is the output image showing the 3 scatterers at the correct ground truth locations, ${x=(-1.93~\mathrm{m}, 1~\mathrm{m}, 0.75~\mathrm{m})}$ and ${y=(2.3~\mathrm{m}, 1.73~\mathrm{m}, -1.3~\mathrm{m})}$.

\begin{figure}[htb]
    \centering
    \subfloat{\includegraphics[width=0.48\columnwidth]{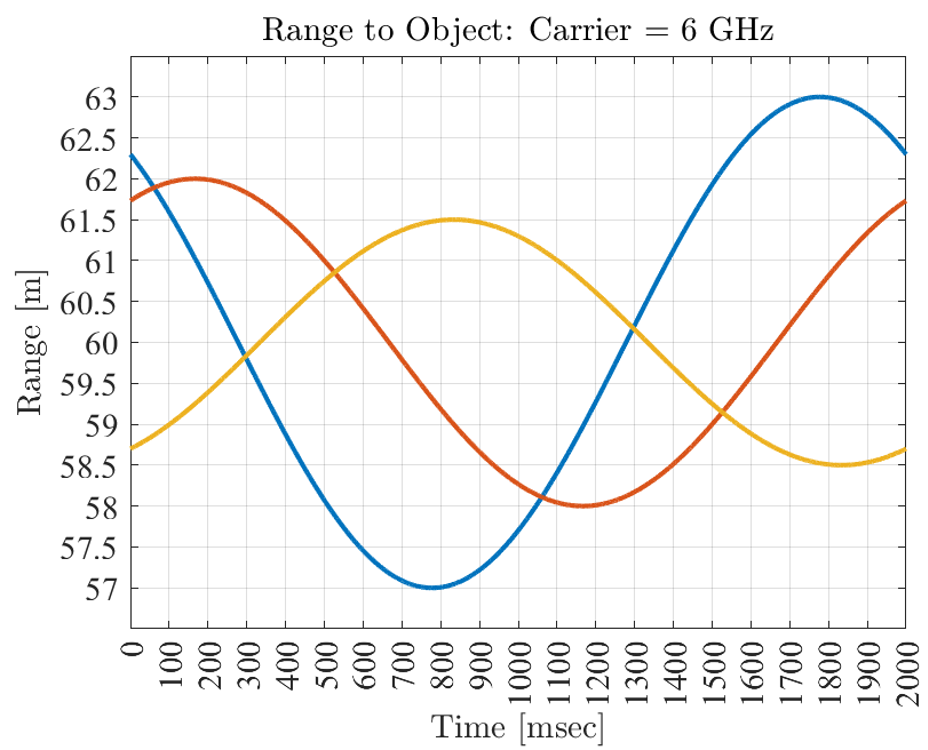}}\hspace{1pt}
    \subfloat{\includegraphics[width=0.48\columnwidth]{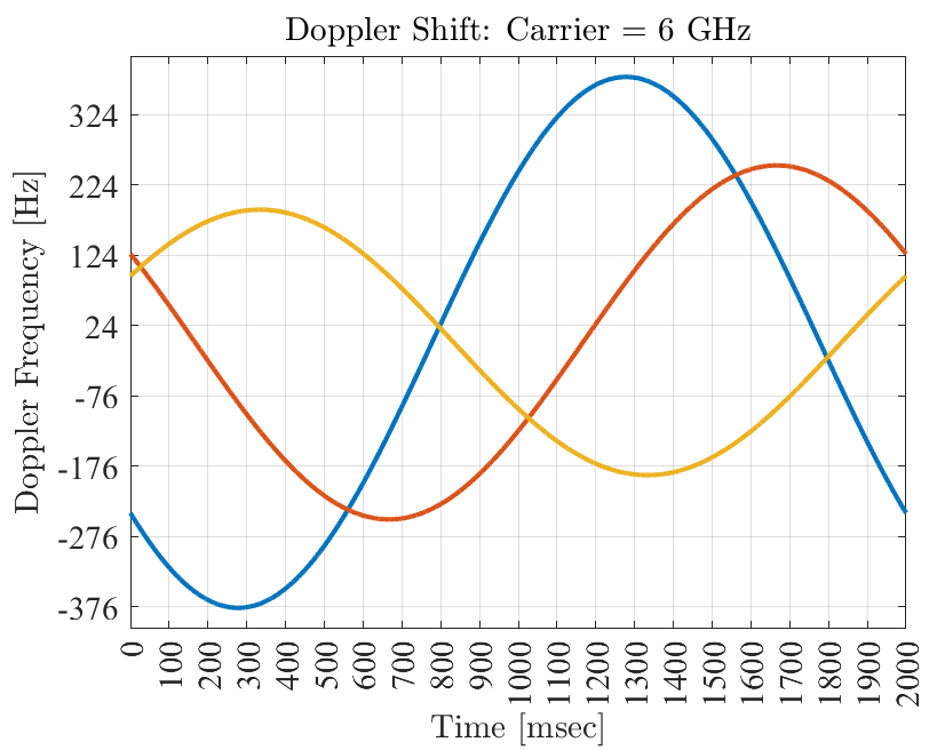}}\hspace{1pt}
    \vspace{-10pt}
	\caption{(Left) Range to scatterers versus time.  (Right) Scatterer Doppler shift versus time.}
\label{fig:Fig-3}
\end{figure}

Fig. \ref{fig:Fig-3} shows the time varying range from the antenna to the scatterers and the Doppler shifts due to rotation.

The next simulation scenario shows the generation of a point null in the image using a continuous phase offset function.  In this example, the carrier frequency is 600~MHz and there is a single scatterer rotating at ${\omega_r = \pi}$ rad/sec.  The initial offsets are ${r_0=1.5~\mathrm{m}}$ and ${\theta_0=300^{\circ}}$.  The distance to the axis of rotation is ${R_a = 60~\mathrm{m}}$ and the ground truth location of the scatterer is at ${x=0.75~\mathrm{m}}$ and ${y=-1.3~\mathrm{m}}$ .  The location of the desired null is at ${x = -1.85~\mathrm{m}}$ and ${y=0.29~\mathrm{m}}$ .  The signal amplitude is 3~V.

\begin{figure}[htb]
    \centering
    \subfloat{\includegraphics[width=0.48\columnwidth]{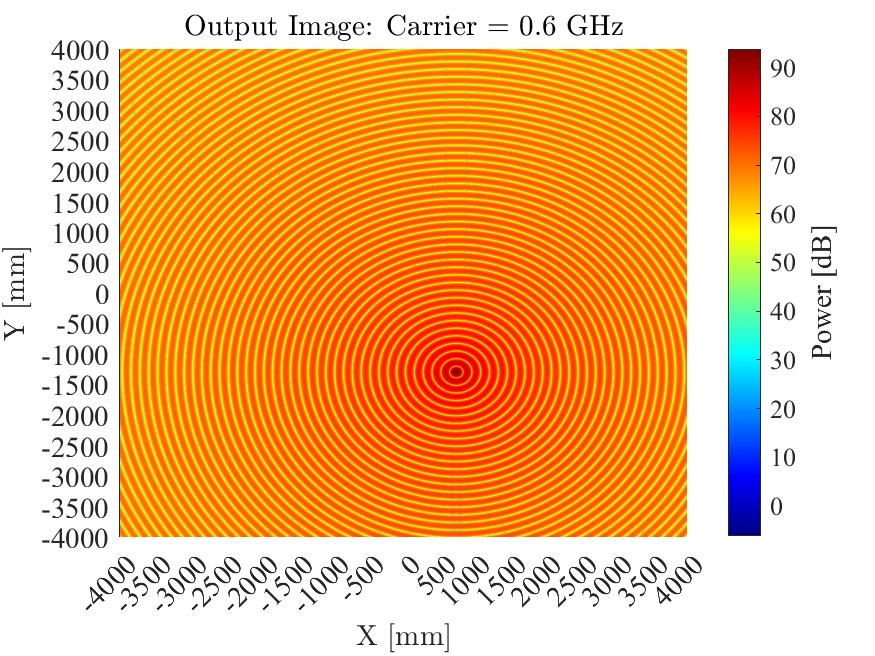}}\hspace{1pt}
    \subfloat{\includegraphics[width=0.48\columnwidth]{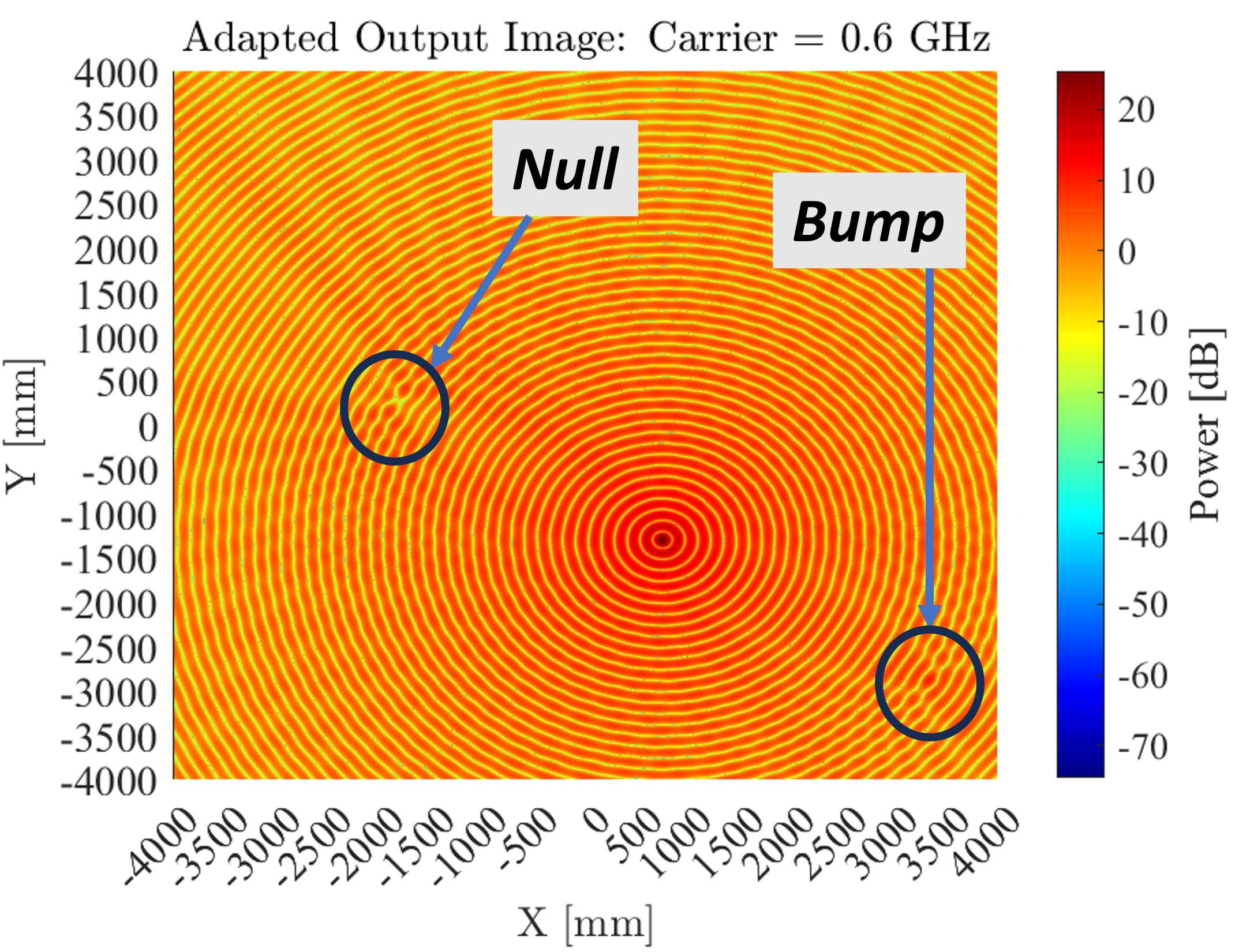}}\hspace{1pt}
    \vspace{-10pt}
	\caption{(Left) Output image without null showing single scatterer.  (Right) Adapted image with desired point null and a corresponding bump in the sidelobe level.}
\label{fig:Fig-4}
\end{figure}

Fig. \ref{fig:Fig-4} shows the adapted image with and without the desired null.  Note that the null creates a corresponding bump in the sidelobe level along a radial line through the signal peak and at a symmetric offset. 

\begin{figure}[htb]
    \centering
    \subfloat{\includegraphics[width=0.48\columnwidth]{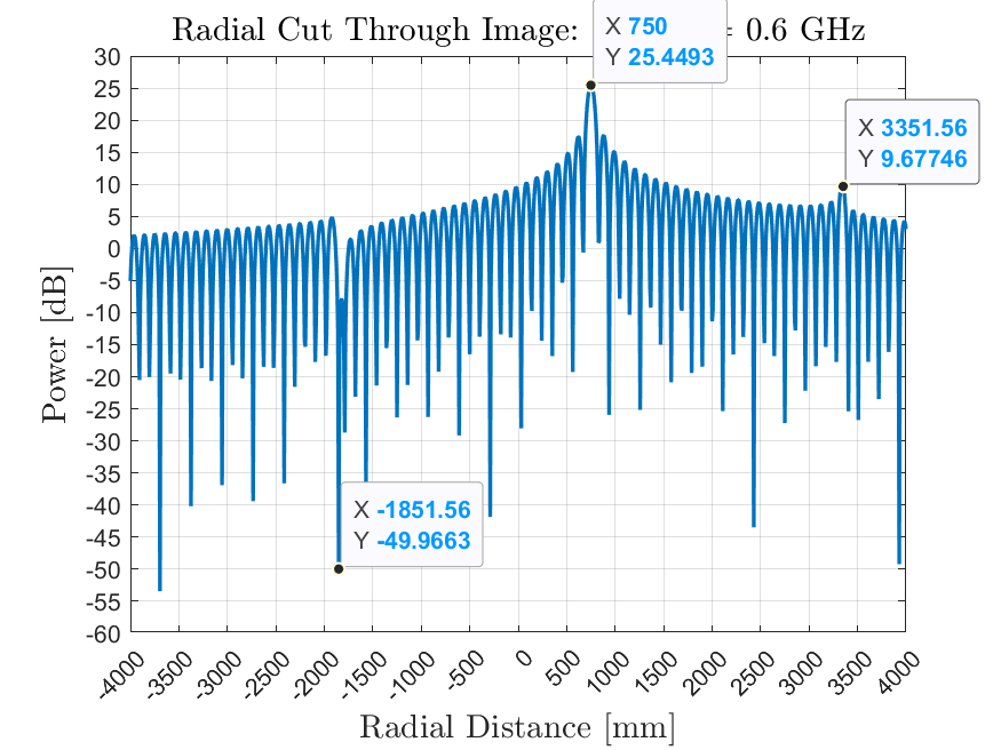}}\hspace{1pt}
    \subfloat{\includegraphics[width=0.48\columnwidth]{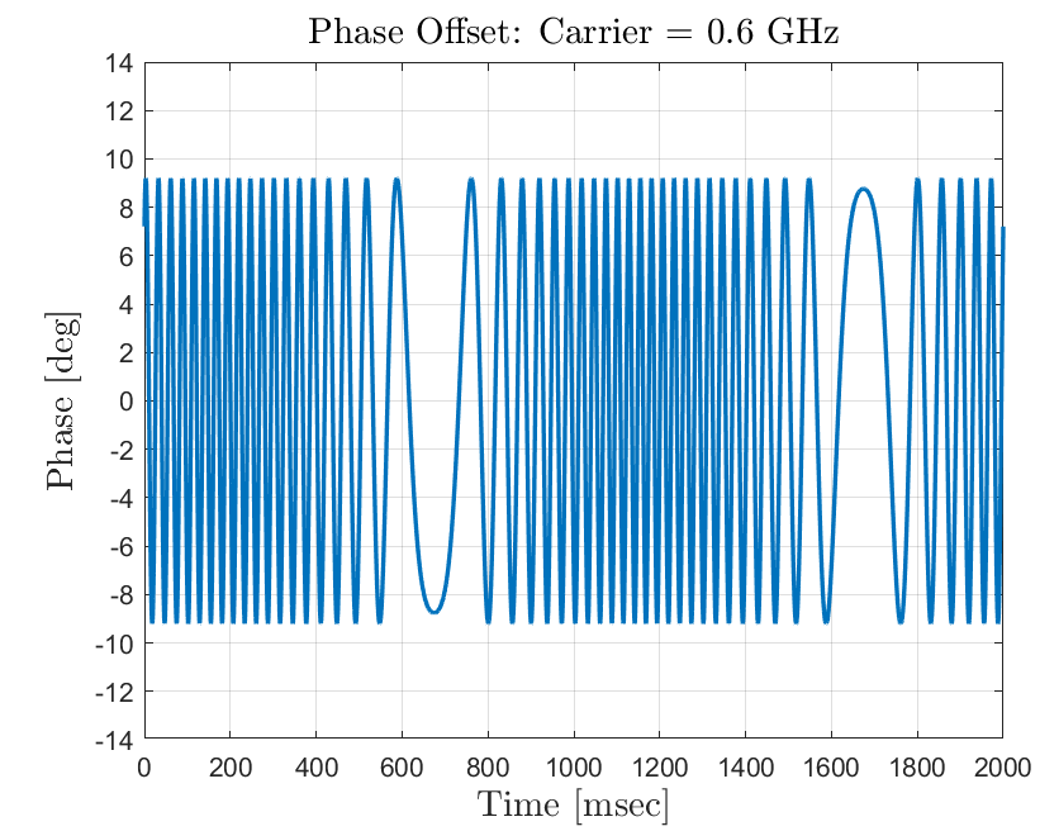}}\hspace{1pt}
    \vspace{-10pt}
	\caption{(Left) Radial cut through image showing the desired null and a bump in the sidelobe level at a symmetric offset from the signal peak.  (Right) Phase offset function used to create the desired point null.}
\label{fig:Fig-5}
\end{figure}

Fig. \ref{fig:Fig-5} is a radial slice through the signal peak which clearly shows the desired null and the corresponding bump in sidelobe level.  Fig. \ref{fig:Fig-5} also shows the phase offset function applied to the received signal to create the null.

\begin{figure}[htb]
    \centering
    \subfloat{\includegraphics[width=0.48\columnwidth]{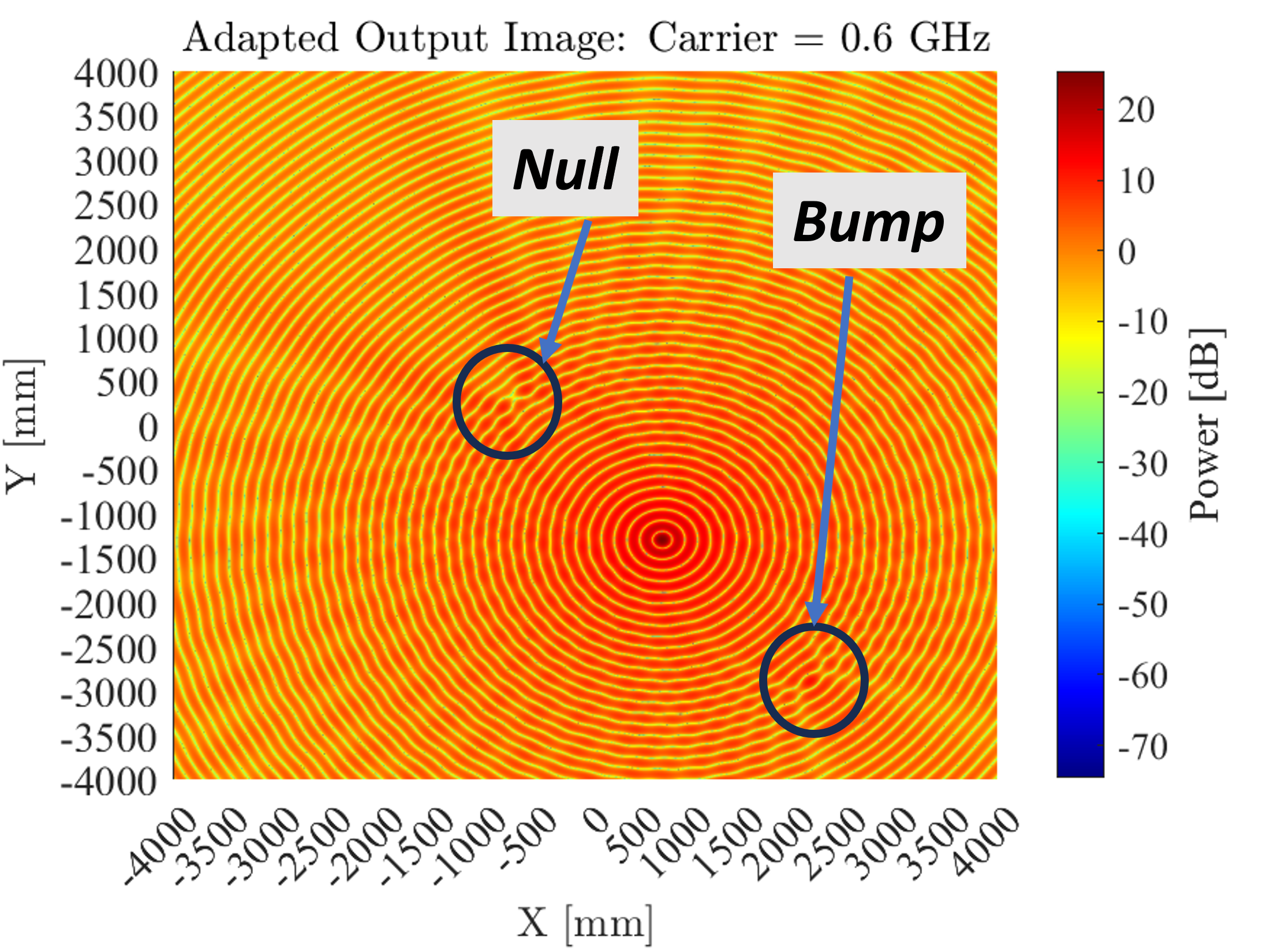}}\hspace{1pt}
    \subfloat{\includegraphics[width=0.48\columnwidth]{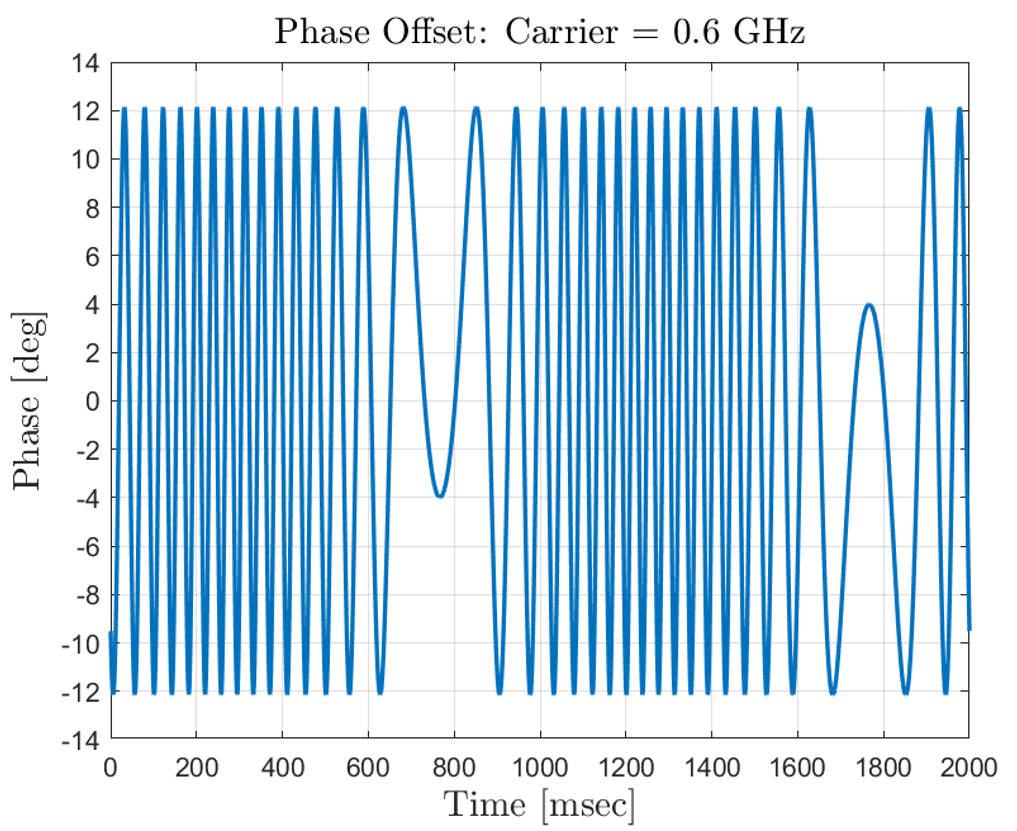}}\hspace{1pt}
    \vspace{-10pt}
	\caption{(Right) Adapted image for a null close to the origin.  (Left) Phase offset shows larger peak-to-peak variation..}
\label{fig:Fig-6}
\end{figure}

\begin{figure}[htb]
    \centering
    \subfloat{\includegraphics[width=0.48\columnwidth]{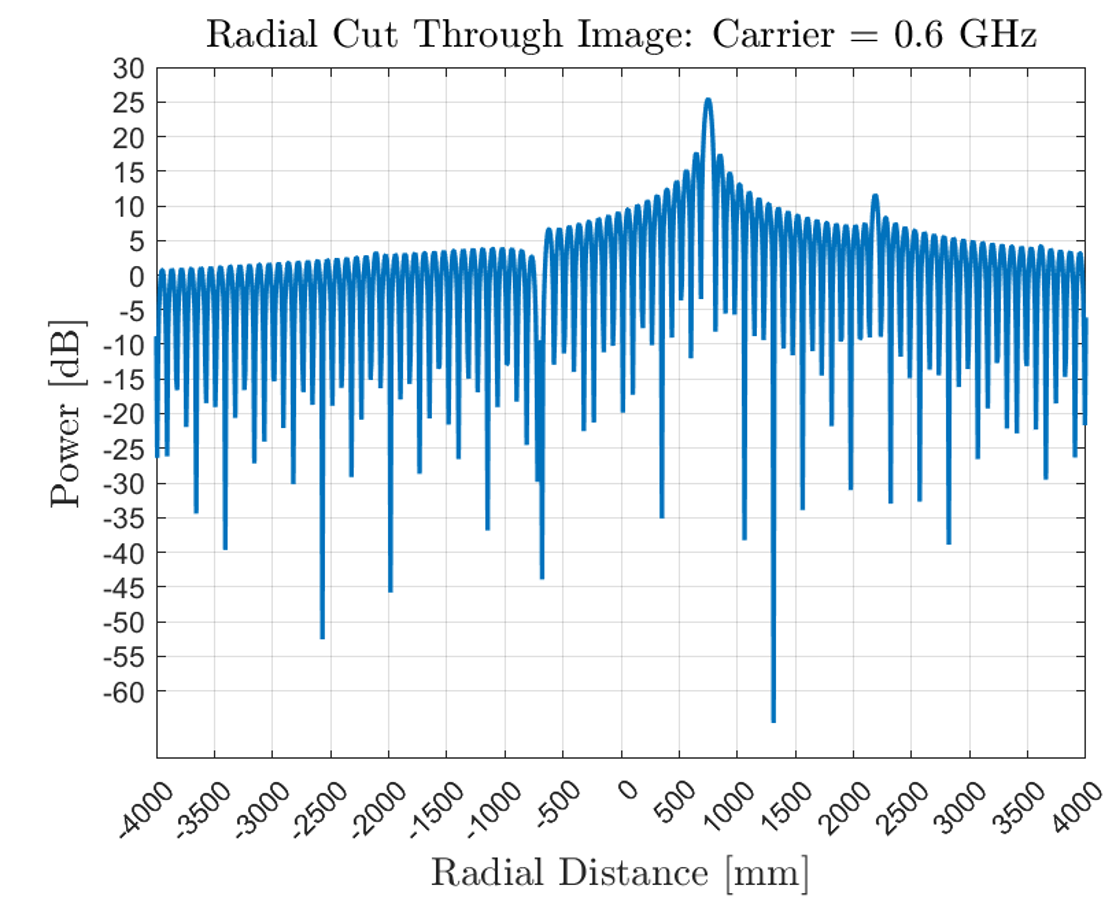}}\hspace{1pt}
    \subfloat{\includegraphics[width=0.48\columnwidth]{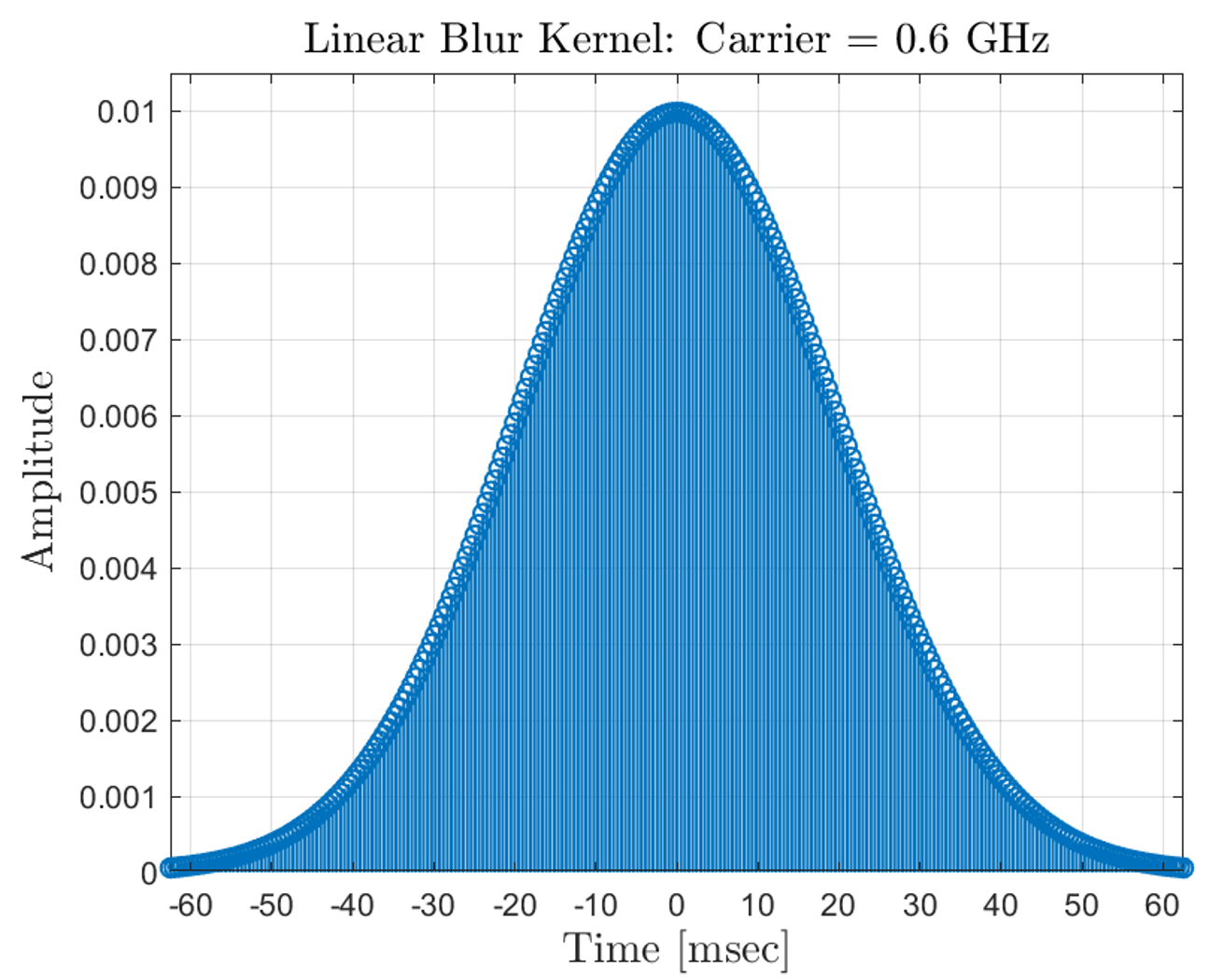}}\hspace{1pt}
    \vspace{-9pt}
	\caption{(Left) Radial cut through the adapted image.  (Right) Linear blurring kernel that represents the Rydberg sensor.}
\label{fig:Fig-7}
\end{figure}

In the next example, the null was placed at a lower spatial frequency, i.e. closer to the origin.  The coordinates of the null are now at ${x = -0.68~\mathrm{m}}$ and ${y=0.29~\mathrm{m}}$.  Fig. \ref{fig:Fig-6} shows the adapted image and the phase offset function used to create the null.  In this case, it is clear that the peak-to-peak excursion of the phase offset function is larger.  Fig. \ref{fig:Fig-7} illustrates a radial cut through the null and the signal peak.
The last simulation example demonstrates the image deblurring approach.  Fig. \ref{fig:Fig-7} illustrates the Gaussian impulse response of a blurring kernel that convolves the received signal.

\begin{figure}[h!]
    \centering
    \subfloat{\includegraphics[width=0.48\columnwidth]{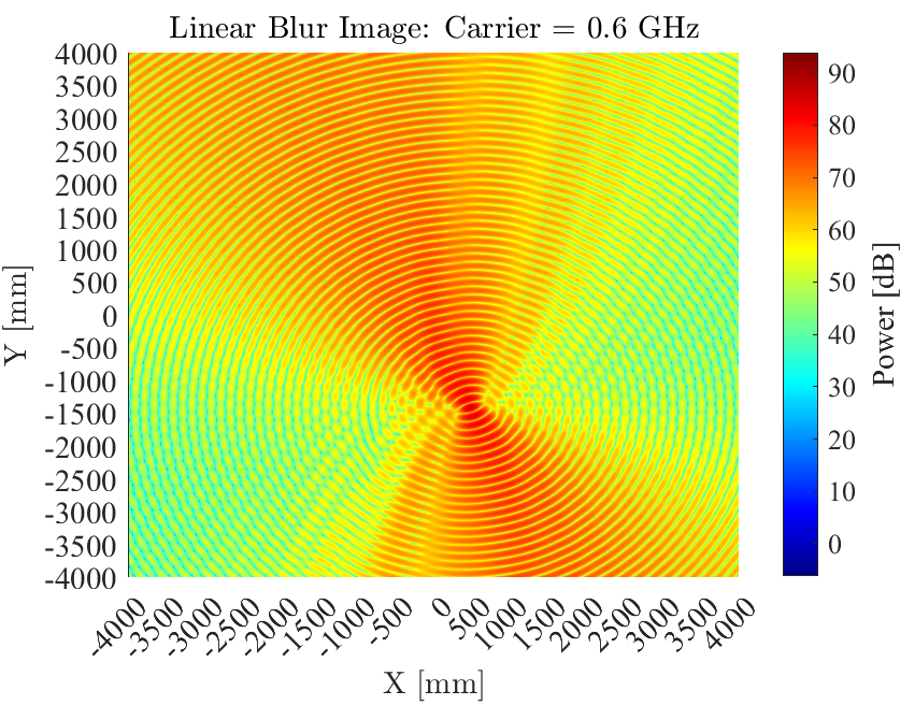}}\label{fig:blurred_image} \hspace{1pt}
    \subfloat{\includegraphics[width=0.48\columnwidth]{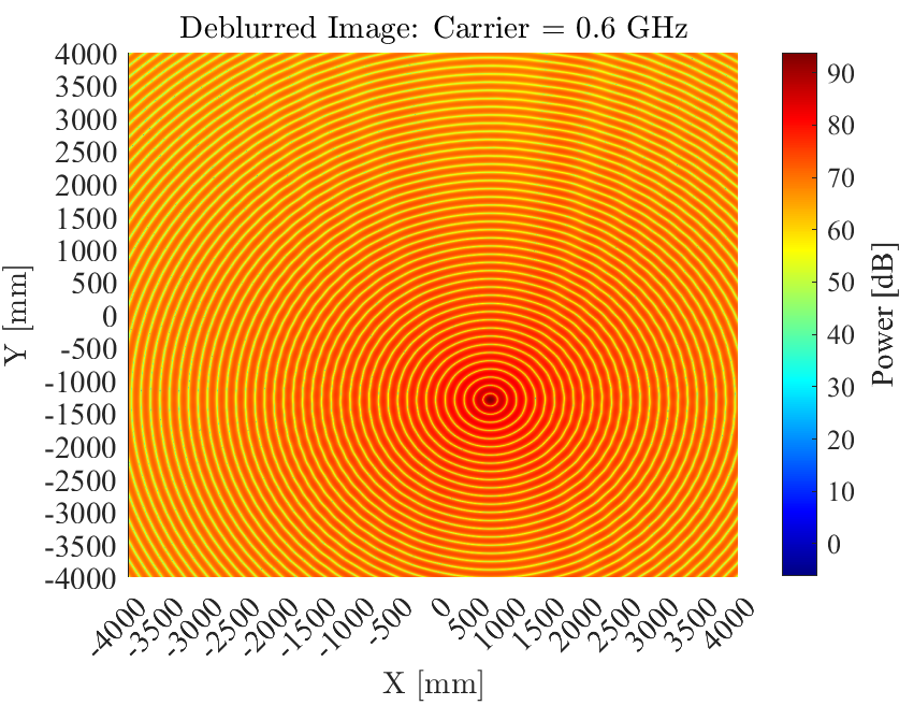}}\label{fig:deblurred_image} \hspace{1pt}
    \vspace{-10pt}
	\caption{(Left) Blurred image due to distortion from the Rydberg sensor.  (Right) Deblurred image after ideal signal is recovered.}
\label{fig:Fig-8}
\vspace{-8pt}
\end{figure}

\begin{figure}[htb]
    \centering
    \subfloat{\includegraphics[width=0.48\columnwidth]{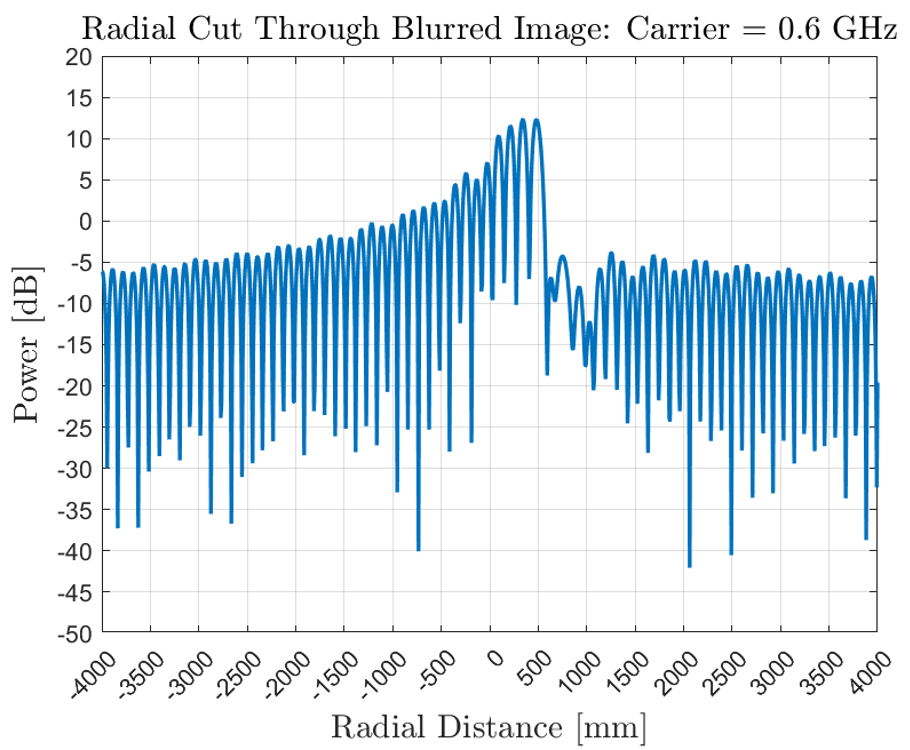}}\hspace{1pt}
    \subfloat{\includegraphics[width=0.48\columnwidth]{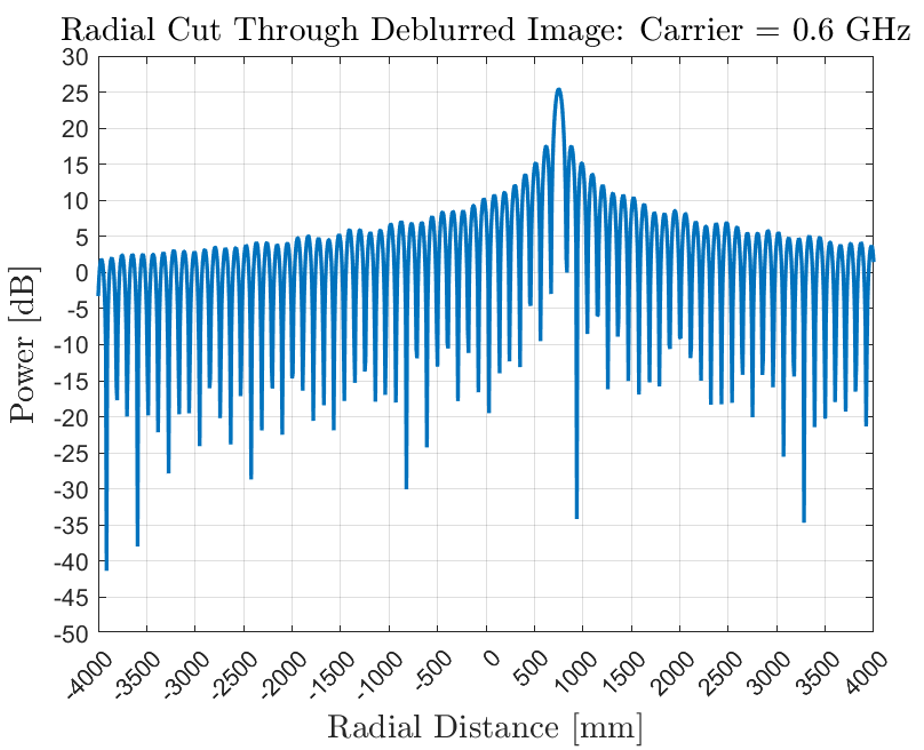}}\hspace{1pt}
    \vspace{-9pt}
	\caption{(Left) Radial cut through blurred image.  (Right) Radial cut through deblurred image.}
\label{fig:Fig-9}
\end{figure}

Fig. \ref{fig:Fig-8} illustrates the blurred image and the result after deblurring is applied to recover the ideal signal.  Fig. \ref{fig:Fig-9} illustrates radial cuts through the signal peak of the blurred and deblurred images.

\vspace{-9pt}

\vspace{-9pt}
\section{Conclusions}
This paper presents a framework for the use of Rydberg quantum sensors in computational imaging applications.  Our contribution provides theoretical results as well as a novel method for placing nulls in a tomographic image by continuously modulating the phase of the radiated LO signal.  Lastly, we present an approach for deblurring the output image if distorted by a linear blurring kernel.

\bibliographystyle{IEEEbib}
\bibliography{strings,refs}

\end{document}